\documentclass[twocolumn,pra,aps,showpacs]{revtex4}

\usepackage{psfrag,graphicx}

\usepackage{dcolumn}
\usepackage{amsmath,amssymb}
\usepackage{bm}
\usepackage[dvips]{color}

\definecolor{mygrey}{gray}{0.35}
\definecolor{mygreen}{rgb}{0.85,1,0.9}
\definecolor{myzard}{cmyk}{0,0,0.05,0}
\definecolor{mywhite}{rgb}{1,1,1}
\definecolor{myred}{rgb}{1,0,0}

 \def\ee{\mathord{\rm e}}
 
 \def\ii{\mathord{\rm i}}

\def\half{\textstyle\frac{1}{2}}

\def\vec#1{{\bf{#1}}} 

\def\bra#1{\langle#1|} \def\ket#1{|#1\rangle}

\begin{document}

\title[Short Title]{
Distillation Protocols for Mixed States of Multilevel Qubits and \\ the
Quantum Renormalization Group}

\author{M.A. Mart\'{\i}n-Delgado and M. Navascu\'es}
\affiliation{
Departamento de F\'{\i}sica Te\'orica I, Universidad Complutense,
28040. Madrid, Spain.
}

\begin{abstract}
We study several properties of distillation protocols to purify
multilevel qubit states (qudits) when applied to a certain family
of initial mixed bipartite states. We find that it is possible
to use qudits states to increase the stability region obtained
with the flow equations to distill qubits. In particular, for
qutrits we get the phase diagram of the distillation process
with a rich structure of fixed points.
We investigate the large-$D$ limit of qudits protocols and find
an analytical solution in the continuum limit.
The general solution of the distillation recursion relations is
presented in an appendix.
We stress the notion of weight amplification for distillation
protocols as opposed to the quantum amplitude amplification that
appears in the Grover algorithm. Likewise, we investigate the
relations between quantum distillation and quantum renormalization
processes.
\end{abstract}

\pacs{03.67.-a, 03.67.Lx}
\maketitle

\section{Introduction}
\label{sec1}
The experimental analysis of the intriguing properties of
entanglement in quantum mechanics requires the availability of   stable
sources of entanglement. Despite the nice properties
exhibited by entanglement, it has the odd behaviour of degrading
by the unavoidable contact with the external environment.
Thus, for the entanglement to be assessed as  a precious mean,
we must devise some method to pump it up to the entanglement source
in order to sustain a prescribed degree of entanglement that we may
need whether for quantum communication protocols (teleportation, cryptography,
dense coding) or quantum computing (algorithmics)
(for a review see \cite{review1}, \cite{review2} and references therein).

Quantum distillation or purification protocols are precisely those methods,
that have been devised to regenerate  entanglement leakages of an
entanglement source. Here we are interested in the purification of
mixed states of bipartite type, having in mind the realization of a
communication protocol by two parties, Alice and Bob. The seminal work
of \cite{bennett1} has provided us with a standard
distillation method that has been the focus to developing more protocols
with the aim at improving its original performances. We shall refer to this
distillation protocol as the BBPSSW protocol. There are feasible
experimental proposals for this type of protocols using polarization
beam splitters (PBS) \cite{zeilinger1}.
Likewise, there  also exist methods for the distillation of pure
states \cite{gisin} that have been implemented experimentally \cite{gisin3}.

In addition to the initial purpose for which the quantum distillation
protocols were devised, they have found another very important application
in connection to the problem of quantum error correction: quantum information
needs to be protected from errors even more than classical information
due to its tendency to become decoherent.
To avoid these errors, one can resort to
the ideas of quantum error correction codes \cite{shor}, \cite{steane}
and fault-tolerant quantum computation \cite{gottesman}, \cite{preskill}.
However, entanglement purification is another alternative
to decoherence which gives a more powerful way of dealing with errors
in quantum communication \cite{bennett2}.

In a typical quantum communication experiment, Alice and Bob are two
spatially separated parties sharing pairs of entangled qubits. The type
of operations allowed on these qubits are denoted as LOCC (Local Operations
and Classical Communication): they comprise local unitary operators
$U_{\rm A}\otimes U_{\rm B}$ on each side, local quantum measurements
and communication of the measurement results through a classical channel.
These local quantum operations will suffer from imperfections producing
local errors. Futhermore, Alice and Bob will also face transmission errors
in their quantum channels due to dissipation and noise.
To overcome these difficulties, they will have to set up an entanglement
purification method. In short, a protocol like the BBPSSW creates a reduced
set of maximally entangled pairs (within a certain accuracy)
out of a larger set of imperfectly entangled pairs: entanglement is created
at the expense of wasting extra pairs.
The degree of purity of a mixed entangled pair is measured in terms of its
fidelity with respect to a maximally entangled pure pair, which is the focus
of the purification protocol.
After the BBPSSW protocol, a new distillation protocol was introduced in
\cite{qpa} by the name of Quantum Privacy Amplification(QPA)
which converges much faster to the desired fidelity
\cite{chiara1}, \cite{chiara2}. Other protocols known as quantum repeaters
\cite{repeaters-pur}, \cite{repeaters-com} allow us to stablish quantum
communication over long distances by avoiding absortion or depolarization
errors that scale exponentially with the length of the quantum channel.

The advantages of dealing with $D$-dimensional or multilevel quantum states
(qudits) instead of qubits are quite apparent: an increase in the information
flux through the communication channels that could speed up
quantum cryptography, etc. \cite{pasqui1}, \cite{pasqui2}, \cite{karlson}.
Thus, it has been quite
natural to propose extensions of the purification protocols for qudits.
One of the proposals \cite{horodeckis} relies on an extension of the
CNOT gate that
is unitary, but not Hermitian. Recently, another very nice proposal has been
introduced \cite{gisin1}, \cite{gisin2}
based on a generalization of the CNOT gate that is both unitary and Hermitian
and  gives a higher convergence.
In this paper we  make a study of the new purification protocols of
\cite{gisin1}, \cite{gisin2} when they are applied to mixed bipartite states
of qudits that are not of the Werner form.
In this way, we combine some of the tools employed by the QPA protocols
\cite{qpa} with the advantages of the new methods.

This paper is organized as follows: in Sect.~\ref{sec2} we review simple
distillation protocols for qubits not in Werner states and we generalize
them for the purification of any of the Bell states. In Sect.~\ref{sec3}
we extend the previous protocols to deal with multilevel qubits and
obtain several results like an improvement in the size of the stability
fidelity basin, analytical formulas for the distillation flows,
phase diagrams, etc. In Sect.~\ref{sec4} we apply the distillation protocols
for the purification of non-diagonal mixed states that are more easily
realized experimentally. In Sect.~\ref{sec5} we study the large-$D$ limit
of these protocols. In Sect.~\ref{sec6} we present a detailed investigation
of the relationships between quantum distillation protocols and
renormalization methods for quantum lattice Hamiltonians.
Section~\ref{sec7} is devoted to conclusions. In appendix~\ref{app}
we find the general solution for the distillation recursion relations used in
the text in the general case of qudits.

\section{Simple Distillation Protocols with Qubits}
\label{sec2}
Our starting point is the orthonormal basis of Bell states formed
by the first qubit belonging to Alice and the second to Bob:
\begin{equation}
\begin{split}
\ket{\Phi^{\pm}} &:= \frac{1}{\sqrt{2}}[\ket{00}\pm \ket{11}]\\
\ket{\Psi^{\pm}} &:= \frac{1}{\sqrt{2}}[\ket{01}\pm \ket{10}]
\end{split}
\label{s1}
\end{equation}
We shall use the word ``simple'' applied to the distillation protocols
to denote that the mixed state we shall be dealing with is made up of
a combination of one state in the set $S:=\{ \ket{\Phi^+}, \ket{\Phi^-}\}$
of Bell states that have coincident bits in
Alice's and Bob's qubits,
with another state in the set $A:=\{ \ket{\Psi^+}, \ket{\Psi^-}\}$
of Bell states that do not have coincidences.
Thus, we have 4 possible combinations to do this type
of entanglement distillation.

To begin with, we shall choose the following mixed state in order to set up
a simple distillation protocol
\begin{equation}
\rho_{++} := F\ket{\Phi^{+}}\bra{\Phi^{+}} +
(1-F)\ket{\Psi^{+}}\bra{\Psi^{+}}
\label{s2}
\end{equation}
Alice and Bob will also need to apply the CNOT gate defined as usual
\begin{equation}
U_{\rm CNOT} \ket{i}\ket{j}:= \ket{i}\ket{i\oplus j}, \ i,j=0,1.
\label{s2b}
\end{equation}

The distillation protocol can be arranged into a set of 5
instructions or steps \cite{bennett1},\cite{qpa},\cite{chiara2}:
\vspace{10 pt}
\begin{center}
\underline{Distillation Protocol for Qubits}
\end{center}
\begin{enumerate}

\item Set up $\rho \longrightarrow \rho \otimes \rho$
with fidelity $F$.

\item Apply bilateral CNOT gate: $U_{\rm BCNOT}$.

\item Alice and Bob measure target qubits.

\item Classical communication of results: retain coincidences
($0_A0_B$ or $1_A1_B$).

\item Go to step 1) with $\rho'$ with fidelity $F'>F$.

\end{enumerate}

\begin{figure}[ht]
\psfrag{A}[Bc][Bc][1][0]{Alice}
\psfrag{B}[Bc][Bc][1][0]{Bob}
\psfrag{F}[Bc][Bc][1][0]{$F$}
\psfrag{G}[Bc][Bc][1][0]{$F$}
\psfrag{H}[Bc][Bc][1][0]{$F$}
\psfrag{I}[Bc][Bc][1][0]{$F$}
\psfrag{C}[Bc][Bc][1][0]{CNOT}
\psfrag{D}[Bc][Bc][1][0]{CNOT}
\psfrag{P}[Bc][Bc][1][0]{Quantum}
\psfrag{Q}[Bc][Bc][1][0]{Distillation}
\psfrag{J}[Bc][Bc][1][0]{$F'>F$}
\psfrag{K}[Bc][Bc][1][0]{$F'>F$}
\includegraphics[width=6 cm]{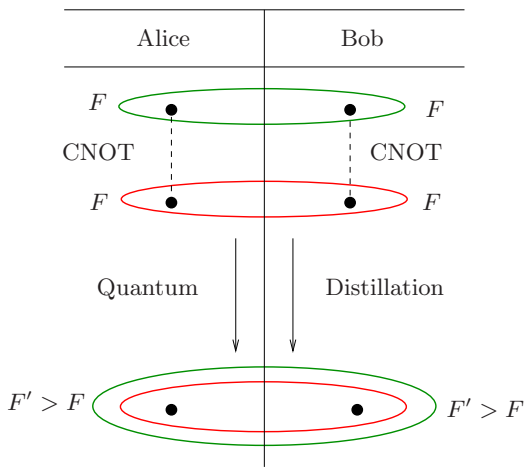}
\caption{Schematic representation of the distillation protocol by
Alice and Bob. Originally, two pairs of shared entangled qubits represented
by enclosed dots are transformed into a single pair of higher purity
(doubly enclosed dots).}
\label{fig1}
\end{figure}

The simplicity of this protocol also relies on the fact that we do not
need any depolarization step, as it is the case when dealing with Werner
states \cite{bennett1}. In Fig.~\ref{fig1} we show a schematic picture
of a single aplication of the purification method.
Let us comment on the outcomes corresponding to the most relevant steps
in this protocol. After step 1/, the 4-quit mixed state $\rho \otimes \rho$
shared by Alice and Bob reads as follows
\begin{equation}
\begin{split}
&\rho_{++} \otimes \rho_{++}
= F^2\ket{\Phi^{+}\Phi^{+}}\bra{\Phi^{+}\Phi^{+}}\\
             &+ F(1-F)[\ket{\Phi^{+}\Psi^{+}}\bra{\Phi^{+}\Psi^{+}} +
\ket{\Psi^{+}\Phi^{+}}\bra{\Psi^{+}\Phi^{+}}] \\
 &+ (1-F)^2\ket{\Psi^{+}\Psi^{+}}\bra{\Psi^{+}\Psi^{+}}
\end{split}
\label{s3}
\end{equation}
In step 2/, Alice and Bob apply bilaterally the CNOT gate taking their
first qubit as source and their second qubit as target, i.e., qubits
first and third are source qubits while qubits second and fourth are
target qubits. To obtain the transformed mixed state we must determine
the action of the bilateral CNOT gate $U_{\rm BCNOT}$ \cite{bennett1}
on the states of the form
$\ket{\varphi_A\varphi_B}$. The results of this computation are shown in
Table~\ref{table1}. With the help of this table we find
\begin{equation}
\begin{split}
& U_{\rm BCNOT} \rho_{++} \otimes \rho_{++} U_{\rm BCNOT} =
F^2\ket{\Phi^{+}\Phi^{+}}\bra{\Phi^{+}\Phi^{+}}\\
             &+ F(1-F)[\ket{\Phi^{+}\Psi^{+}}\bra{\Phi^{+}\Psi^{+}} +
\ket{\Psi^{+}\Psi^{+}}\bra{\Psi^{+}\Psi^{+}}] \\
 &+ (1-F)^2\ket{\Psi^{+}\Phi^{+}}\bra{\Psi^{+}\Phi^{+}}
\end{split}
\label{s4}
\end{equation}

\begin{figure}[ht]
\psfrag{0}[Bc][Bc][1][0]{0}
\psfrag{1}[Bc][Bc][1][0]{1}
\psfrag{2}[Bc][Bc][1][0]{$F_c$}
\psfrag{c}[Bc][Bc][1][0]{$\half$}
\psfrag{g}[Bc][Bc][1][0]{$F$}
\includegraphics[width=6 cm]{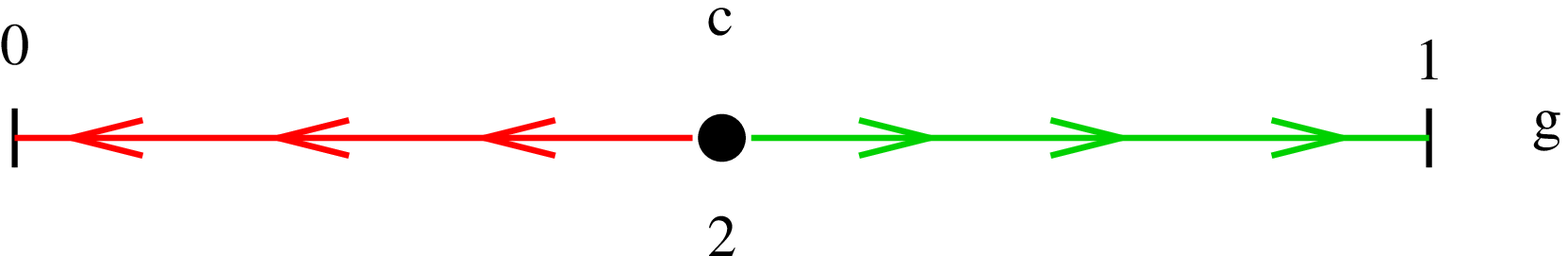}
\caption{Flow diagram for the fidelity $F$ of the distillation protocol
given by the recursion relation (\ref{s6}).}
\label{flowdiagram}
\end{figure}

After steps 3/ and 4/, Alice and Bob measure their target qubits and retain
their source qubit whenever they find, via classical communication, the same
results: either $0_A0_B$ or $1_A1_B$. This fact selects the state
$\ket{\Phi^+}$ as the only admissible possibility for the target state.
Thus, only the first and third terms in the RHS of (\ref{s4}) survive to
this process and the resulting 2-qubit state $\rho'_{++}$
is again of the same form as
the original starting state (\ref{s2}) in step 1/,
but with a higher fidelity $F'>F$. In fact, we get
\begin{equation}
\rho'_{++} := F'\ket{\Phi^{+}}\bra{\Phi^{+}} +
(1-F')\ket{\Psi^{+}}\bra{\Psi^{+}}
\label{s5}
\end{equation}
with the new fidelity being
\begin{equation}
F' = \frac{F^2}{F^2 + (1-F)^2}
\label{s6}
\end{equation}
This relation defines a recursion scheme for entanglement purification:
starting with say $N_P$ pairs of Bell states of fidelity $F$, after
every application of the whole protocol we obtain $\frac{N_P}{2}$ pairs
of higher fidelity $F'>F$.
Thus, purification is achieved at the expense of halving the number of
Bell pairs. The fixed points $F_c$ of the recursion relation (\ref{s6}) are
defined as $F'(F_c):=F_c$ and they are given by $F_c=0,\half,1$.
The fixed points $F_c=0,1$ are stable, while $F_c=\half$ is unstable.
The best way to recast these qualitative properties of the flow equation
for the fidelities (\ref{s6}) is to draw the corresponding  flow diagram
as shown in Fig.~\ref{flowdiagram}.
\begin{table}[t]
\begin{ruledtabular}
\begin{tabular}{cc}
$\ket{\varphi_A}\ket{\varphi_B}$ &
$U_{\rm BCNOT}\ket{\varphi_A}\ket{\varphi_B}$\\
 \hline \hline
$\ket{\Phi^{+}}\ket{\Phi^{+}}$ &  $\ket{\Phi^{+}}\ket{\Phi^{+}}$ \\
\hline
$\ket{\Phi^{+}}\ket{\Psi^{+}}$ &  $\ket{\Phi^{+}}\ket{\Psi^{+}}$ \\
\hline
$\ket{\Psi^{+}}\ket{\Phi^{+}}$ &  $\ket{\Psi^{+}}\ket{\Psi^{+}}$ \\
\hline
$\ket{\Psi^{+}}\ket{\Psi^{+}}$ &  $\ket{\Psi^{+}}\ket{\Phi^{+}}$ \\
\end{tabular}
\end{ruledtabular}
\caption{This table shows the results of applying the bilateral CNOT
gate to certain pairs of Bell states needed to distillation.}
\label{table1}
\end{table}

Next, we may wonder whether it is possible to devise distillation protocols
for the three possible combinations of Bell states, namely,
\begin{equation}
\begin{split}
\rho_{+-} &:= F\ket{\Phi^{+}}\bra{\Phi^{+}} +
(1-F)\ket{\Psi^{-}}\bra{\Psi^{-}} \\
\rho_{-+} &:= F\ket{\Phi^{-}}\bra{\Phi^{-}} +
(1-F)\ket{\Psi^{+}}\bra{\Psi^{+}} \\
\rho_{--} &:= F\ket{\Phi^{-}}\bra{\Phi^{-}} +
(1-F)\ket{\Psi^{-}}\bra{\Psi^{-}}
\end{split}
\label{s7}
\end{equation}
We can answer this question affirmatively by computing the action
of the bilateral CNOT gate on the tensor product of
these mixed states (\ref{s7}). With a similar analysis which has led us
to Table~\ref{table1} \cite{bennett1}, we obtain
\begin{equation}
\begin{split}
& U_{\rm BCNOT} \rho_{+-} \otimes \rho_{+-} U_{\rm BCNOT} =
F^2\ket{\Phi^{+}\Phi^{+}}\bra{\Phi^{+}\Phi^{+}}\\
             &+ F(1-F)[\ket{\Phi^{-}\Psi^{-}}\bra{\Phi^{-}\Psi^{-}} +
\ket{\Psi^{-}\Psi^{+}}\bra{\Psi^{-}\Psi^{+}}] \\
 &+ (1-F)^2\ket{\Psi^{+}\Phi^{-}}\bra{\Psi^{+}\Phi^{-}}\\
& U_{\rm BCNOT} \rho_{-+} \otimes \rho_{-+} U_{\rm BCNOT} =
F^2\ket{\Phi^{+}\Phi^{-}}\bra{\Phi^{+}\Phi^{-}}\\
             &+ F(1-F)[\ket{\Phi^{-}\Psi^{+}}\bra{\Phi^{-}\Psi^{+}} +
\ket{\Psi^{-}\Psi^{-}}\bra{\Psi^{-}\Psi^{-}}] \\
 &+ (1-F)^2\ket{\Psi^{+}\Phi^{+}}\bra{\Psi^{+}\Phi^{+}} \\
& U_{\rm BCNOT} \rho_{--} \otimes \rho_{--} U_{\rm BCNOT} =
F^2\ket{\Phi^{+}\Phi^{-}}\bra{\Phi^{+}\Phi^{-}}\\
             &+ F(1-F)[\ket{\Phi^{+}\Psi^{-}}\bra{\Phi^{+}\Psi^{-}} +
\ket{\Psi^{+}\Psi^{-}}\bra{\Psi^{+}\Psi^{-}}] \\
 &+ (1-F)^2\ket{\Psi^{+}\Phi^{-}}\bra{\Psi^{+}\Phi^{-}}
\end{split}
\label{s8}
\end{equation}
We now realize that if we proceed to measure the target bits and classical
communicate the results, we do not end up with the same type of mixed state
as we had started with. That is, the protocol as it stands is not valid since
it does not yield invariant mixed states.
This problem has a solution provided we introduce an
additional step prior to the measurement of the target qubits by Alice and Bob.
This additional step corresponds to a local unitary operation
$U_A\otimes U_B$ performed by Alice and Bob on their source qubits.
The form of this local unitary depends on the mixed state we are distilling.
We find the following results:

\noindent Step $2'$. Alice and Bob apply a local unitary transformation
$U_A\otimes U_B$ to their source qubits:

\noindent For $\rho_{+-}$,
$U_A = \half (1+\ii)(\sigma_x+\sigma_y)$,
$U_B = \half (1-\ii)(\sigma_x-\sigma_y)$.

\noindent For $\rho_{-+}$,
$U_A = U_B = \half (1+\ii)(\sigma_x+\sigma_y)$.

\noindent For $\rho_{--}$,
$U_A = \sigma_z$, $U_B = 1$.

After this extra step, we can guarantee that the resulting 4-qubit
mixed state has the appropriate Bell pairs at the source qubits so
as to produce the same original state, once steps 3/ and 4/ are performed.
Moreover, it is straightforward to prove that the new fidelity for these
3 protocols is also given by the same recursion relation (\ref{s6}) as in
the first protocol.

Finally, if the constraint that the state of fidelity $F$ must be
a $\Phi^{\pm}$ state is relaxed, then there are two additional possible
mixed states whose analysis can be carried out in a similar fashion.

\section{Multilevel Extensions of Distillation Protocols}
\label{sec3}
In order to generalize the simple distillation protocol of the
previous section to the case of qudits, we must notice that the two
main ingredients in that distillation protocol are:

\noindent i) the CNOT gate,

\noindent ii) the Bell states (\ref{s1}).

Regarding the CNOT gate,
the extension of this gate to deal with qudits
is not unique.
As has been noted in \cite{gisin1},\cite{gisin2},
the CNOT gate for qubits (\ref{s2b})
has 3 properties that make it special, namely
\begin{equation}
\begin{split}
U_{\rm CNOT}^{\dag} = U_{\rm CNOT}^{-1}, \\
U_{\rm CNOT}^{\dag} = U_{\rm CNOT}, \\
i\oplus j = 0 \Leftrightarrow i=j.
\end{split}
\label{g1}
\end{equation}
The extension of the CNOT gate for qudits that satisfies these 3 properties
(\ref{g1}) is given by \cite{gisin1},\cite{gisin2}
\begin{equation}
U_{\rm CNOT} \ket{i}\ket{j}:=\ket{i}\ket{i\ominus j}, \ i,j=0,\ldots, D-1
\label{g2}
\end{equation}
where $i\ominus j:=i-j, {\rm mod} \ D$, denotes substraction modulus $D$.
This is the definition that we shall adopt throughout this paper, unless
otherwise stated.

As for the higher-dimensional extension of Bell states (\ref{s1}), we shall
also take the following generalization \cite{gisin1},\cite{gisin2}
\begin{equation}
\ket{\Psi_{kj}}:=U_{\rm CNOT} \left[(U_{\rm F}\ket{k})\otimes \ket{j}\right], \ k,j=0,\ldots, D-1
\label{g3}
\end{equation}
where $U_{\rm F}$ is the quantum Fourier transform (QFT)
\begin{equation}
U_{\rm F}\ket{k}:=
\frac{1}{\sqrt{D}}\sum_{y=0}^{D-1}\ee^{\frac{2\pi \ii k y}{D}}\ket{y},
\label{g4}
\end{equation}
which reduces to the Hadamard gate when dealing with qubits ($D=2$).
As a matter of fact, we can readily check that for the special case of qubits
$D=2$ we recover the standard Bell pairs (\ref{s1}) in the following form
\begin{equation}
\begin{split}
\ket{\Psi_{00}} = \ket{\Phi^+}, & \ket{\Psi_{01}} = \ket{\Psi^+},\\
\ket{\Psi_{10}} = \ket{\Phi^-}, & \ket{\Psi_{11}} = \ket{\Psi^-}
\end{split}
\label{g5}
\end{equation}
Moreover, using the generalized CNOT gate (\ref{g2}),
the generalized Bell states are given by
\begin{equation}
\ket{\Psi_{kj}} =
\frac{1}{\sqrt{D}}\sum_{y=0}^{D-1}\ee^{\frac{2\pi \ii k y}{D}}
\ket{y}\ket{y\ominus j}.
\label{g6}
\end{equation}

With these extensions of the CNOT gate and the Bell states, we can set up
a generalization of the simple distillation protocols of Sect. II for
qudit states. These protocols have the same 5 steps as before.

\noindent  {\em Step 1/}. We shall assume  a general diagonal mixed state of the form
\begin{equation}
\begin{split}
\rho &:= \sum_{k,j=0}^{D-1} q_{kj}\ket{\Psi_{ij}}\bra{\Psi_{ij}},\\
1 &=: \sum_{k,j=0}^{D-1} q_{kj},
\end{split}
\label{g7}
\end{equation}
where $q_{kj}$ are normalized probabilities. For non-diagonal mixed states,
we refer to Sect. IV. Then, Alice and Bob share pairs $\rho \times \rho$
of these states (\ref{g7}).

\noindent  {\em Step 2/}. Alice and Bob apply bilaterally the generalized
CNOT gate (\ref{g2}). To know the result of this operation on the state
(\ref{g7}) we need a previous result about the action of the gate
$U_{\rm BCNOT}$ on pairs of generalized Bell states (\ref{g3}).
After some algebra, we arrive at the following expression
\begin{equation}
U_{\rm BCNOT} \ket{\Psi_{kj}}\ket{\Psi_{k'j'}} =
\ket{\Psi_{k\oplus k',j}}\ket{\Psi_{D\ominus k',j\ominus j'}}
\label{g8}
\end{equation}
This is a fundamental result for it means that {\em the space of two-pairs of
generalized Bell states is invariant under the action of the generalized
bilateral CNOT gate.} This is a very nice result that condenses in a
single formula all the possibilities for the outcome of the action of the
CNOT gates on Bell states, in particular, the whole table employed by
Bennett et al. in \cite{bennett1} for the case of qubits is contained in
equation (\ref{g8}).
This property is essential in order to have a closed
distillation protocol.
Actually, it would have been enough to have obtained only the source qubits
as generalized Bell states.

Then, with the help of this property (\ref{g8}) we  obtain the action of
$U_{\rm BCNOT}$ on pairs of $\rho$ states, as follows
\begin{equation}
\begin{split}
& U_{\rm BCNOT} \rho \otimes \rho  U_{\rm BCNOT} =
\sum_{k,j=0}^{D-1} \sum_{k',j'=0}^{D-1}
q_{k\ominus k',j} q_{k'j'} \\
& \ket{\Psi_{kj}\Psi_{D\ominus k',j\ominus j'}}
\bra{\Psi_{kj}\Psi_{D\ominus k',j\ominus j'}}
\end{split}
\label{g9}
\end{equation}
We see that this state is already of the same form in the source qubits as
the original $\rho$ (\ref{g7}).

\noindent  {\em Step 3/}. Alice and Bob measure their target qubits in
(\ref{g9}). To see the result of this measurement, let us write the explicit
form of the target qubits, namely
\begin{equation}
\Psi_{D\ominus k',j\ominus j'} =
\frac{1}{\sqrt{D}}\sum_{z=0}^{D-1}\ee^{\frac{-2\pi \ii k' z}{D}}
\ket{z}\ket{z\ominus (j\ominus j')}.
\label{g10}
\end{equation}
Therefore, coincidences between Alice's and Bob's target qubits will happen
only when the following condition is satisfied
\begin{equation}
z = z\ominus (j\ominus j') \Longleftrightarrow j = j'.
\label{g11}
\end{equation}

\noindent  {\em Step 4/}. After their measurement, Alice and Bob communicate
classically their result so that they retain the resulting source Bell pairs
only when they have coincidences, and discard them otherwise. The resulting
net effect of this process is to produce a Kronecker delta function
$\delta_{jj'}$ in the target qubits. More precisely, the resulting unnormalized
mixed state is given by
\begin{equation}
\rho' \sim \sum_{k,j=0}^{D-1} \sum_{k',j'=0}^{D-1}
q_{k\ominus k',j} q_{k'j'} \delta_{jj'}
\ket{\Psi_{kj}} \bra{\Psi_{kj}}.
\label{g12}
\end{equation}
Therefore, we end up with a diagonal mixed state of the same form as the
starting one
\begin{equation}
\rho' = \sum_{k,j=0}^{D-1}
q'_{kj} \ket{\Psi_{kj}} \bra{\Psi_{kj}}.
\label{g13}
\end{equation}
with the new probabilities given by
\begin{equation}
q'_{kj} = \frac{\sum_{k'=0}^{D-1}q_{k\ominus k',j} q_{k'j}}
{\sum_{k,j=0}^{D-1}\sum_{k'=0}^{D-1}q_{k\ominus k',j} q_{k'j}}.
\label{g14}
\end{equation}
This is a generalized recursion relation that includes eq. (\ref{s6})
as a particular instance.

\noindent  {\em Step 5/}. Alice and Bob start all over again the same process
with the initial state now being $\rho'$ in (\ref{g13}-\ref{g14}).

The nice feature of these generalized distillation protocols for dealing with
qudits is the fact that we have at our disposal explicit analytical formulas
(\ref{g14})
for the evolution (flow) of the different weights (probabilities) of the
generalized mixed states to be purified.
As these distillation protocols are too general, it is worthwhile to consider
some particular cases of interest separately.
The general solution to the distillation recursion relations (\ref{g14})
is presented in the appendix~\ref{app}.
We hereby provide the following
analysis of some examples:

{\em i).} Let us investigate the closest generalization of the
simple protocols introduced in Sect. II. Thus, let us consider the following
type of initial mixed state
\begin{equation}
\begin{split}
\rho &:= \sum_{i=0}^{M} q_i \ket{\Psi_{0i}}\bra{\Psi_{0i}}, \ M\leq D-1,\\
1 &=: \sum_{i=0}^{M}q_i, \ q_i \geq 0,
\end{split}
\label{g15}
\end{equation}
This corresponds to working with the subset of all possible generalized Bell
states of the form $\{ \ket{\Psi_{0i}} \}_{i=0}^{D-1}$. Interestingly enough,
this includes the case of the state $\rho_{++}$ in (\ref{s2}) for $D=2$.
The recursion relations (\ref{g14}) for this special subset of states takes the
following simpler form
\begin{equation}
q'_i = \frac{q_i^2}{\sum_{j=0}^M q_j^2}.
\label{g16}
\end{equation}

 For $M=2$, i.e., considering a mixed state formed of just two Bell states
of the form $\ket{\Psi_{0i}}$, the protocol has the following recursion
relation
\begin{equation}
q'_i = \frac{q_i^2}{q_i^2 + (1-q_i)^2},
\label{g17}
\end{equation}
where here the index $i$ stands for any possible pair of Bell states of the
type $\ket{\Psi_{0i}}$. In other words, we have found a direct $D$-dimensional
generalization of the distillation protocols for qubits in Sect. II,
with $q_i:=F$.

{\em ii).}  For $M=D-1$ and taking $q_0:=F$ and 
$q_i:=\frac{1-F}{D-1}, i=1,\ldots,D-1$
we can find a more advantageous protocol than the previous one.
In fact, in this case we find that
\begin{equation}
q'_0:=F'= \frac{F^2}{F^2 + \frac{(1-F)^2}{D-1}}.
\label{g18}
\end{equation}
The fixed points of this recursion relation are now given by
$F_c=0,\frac{1}{D},1$. Despite being a non-linear recursion relation,
(\ref{g18}) admits an explicit analytical solution for the general
term of the series $F_{k}$ given by
\begin{equation}
F_{k} = \frac{F^{(2^k)}}{F^{(2^k)} +
(D-1) \left[\frac{1-F}{D-1}\right]^{(2^k)}}, \ k\geq 1, F_0:=F.
\label{g18b}
\end{equation}
From this solution, we inmediately find that the fixed points $F_c=0,1$
are stable while $F_c=\frac{1}{D}$ is unstable.

\begin{figure}[t]
\psfrag{x}[Bc][Bc][1][0]{$F$}
\psfrag{y}[Bc][Bc][1][0]{$F'$}
\vspace{5 pt}
\includegraphics[scale = 0.5]{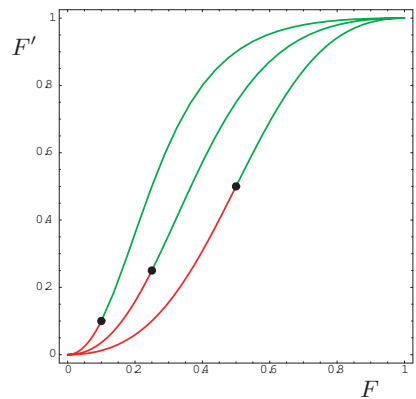}
\caption{Plots of the distilled fidelities $F'$ as a function of the
original fidelity $F$ for several values of the dimension $D$ of the
qudits: $D=2,4,10$.}
\label{recursionFD}
\end{figure}

\begin{figure}[ht]
\psfrag{0}[Bc][Bc][1][0]{0}
\psfrag{1}[Bc][Bc][1][0]{1}
\psfrag{2}[Bc][Bc][1][0]{$F_c$}
\psfrag{c}[Bc][Bc][1][0]{$\frac{1}{D}$}
\psfrag{g}[Bc][Bc][1][0]{$F$}
\includegraphics[width=6 cm]{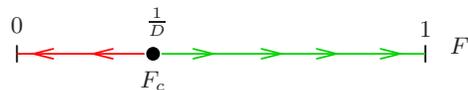}
\caption{Flow diagram for the fidelity $F$ of the generalized
distillation protocol for qudits
given by the recursion relation (\ref{g18}).}
\label{flowdiagramD}
\end{figure}

In Fig.~\ref{recursionFD} we plot the function
$F'=F'(F)$
for several values of the dimension $D$. From the analysis of these curves
we inmediately obtain the corresponding flow diagram that we represent in
Fig.~\ref{flowdiagramD}. We check that for $D=2$ we recover the flow diagram
corresponding to standard qubits (Fig.~\ref{flowdiagram}).

We see from Fig.~\ref{flowdiagramD} that the stability basin is increased
with respect to the case of standard qubits, as in Fig.~\ref{flowdiagram}.
This means that we can start with a mixed state having a fidelity $F$
with respect to the Bell state $\ket{\Psi_{00}}$ lower that $\half$ and
we still will succeed in purifiying that state towards fidelity close to 1.
Thus, we have found that it is more advantageous to distill a given
Bell state $\ket{\Psi_{00}}$ if we prepare the mixed state $\rho$ in
(\ref{g15}) in the form
\begin{equation}
\rho := F \ket{\Psi_{00}} \bra{\Psi_{00}} +
\frac{1-F}{D-1} \sum_{i=1}^{D-1} \ket{\Psi_{0i}} \bra{\Psi_{0i}},
\label{g19}
\end{equation}
rather than using just one single of those states
\begin{equation}
\rho := F \ket{\Psi_{00}} \bra{\Psi_{00}} +
(1-F) \ket{\Psi_{0i}} \bra{\Psi_{0i}}, \ i\neq 0.
\label{g20}
\end{equation}

We may wonder how is it likely for Alice and Bob to obtain the same values
(coincidences) after measuring the target qudits in the step 3/ of the
distillation protocol.
Let us denote by ${\cal P}_{\rm AB}$
this probability which will depend on the value $F$ of the fidelity.
From equations (\ref{g12}) and (\ref{g18}) we find this probability of
coincidences to be
\begin{equation}
{\cal P}_{\rm AB}(F) = F^2 + \frac{(1-F)^2}{D-1}.
\label{g20b}
\end{equation}
The minimum of this probability is at $F_0=\frac{1}{D}$ and its value
is ${\cal P}_{\rm AB}(\frac{1}{D})=\frac{1}{D}$. Likewise,
${\cal P}_{\rm AB}(1)=1$. Thus, we find that the probability is lower
and upper bounded as
${\frac{1}{D}\leq\cal P}_{\rm AB}(F)\leq F$ for $F\in[\frac{1}{D},1]$.

One is also interested in knowing the number of steps $K(\epsilon, F_0)$
needed to achieve
a certain final fidelity close to 1, say $1-\epsilon$, starting from an
appropriate initial fidelity $F_0>\frac{1}{D}$. This number can be computed
from our analytical solution (\ref{g18b}) from the condition
\begin{equation}
F_{K(\epsilon, F_0)}:=1-\epsilon.
\label{g21}
\end{equation}
Thus, we find the following analytical formula for the number of steps
needed to obtain a certain degree of fidelity $\epsilon$ as a function of
the initial fidelity $F_0>\frac{1}{D}$,i.e.,
\begin{equation}
K(\epsilon, F_0) = \left\lceil \log_2 \left(
\frac{\ln (\frac{\epsilon}{(1-\epsilon)(D-1)})}
{\ln (\frac{1-F_0}{(D-1)F_0})}  \right) \right\rceil .
\label{g22}
\end{equation}
\begin{figure}[ht]
\psfrag{x}[Bc][Bc][1][0]{$F_0$}
\psfrag{y}[Bc][Bc][1][0]{$K$}
\includegraphics[width=8 cm]{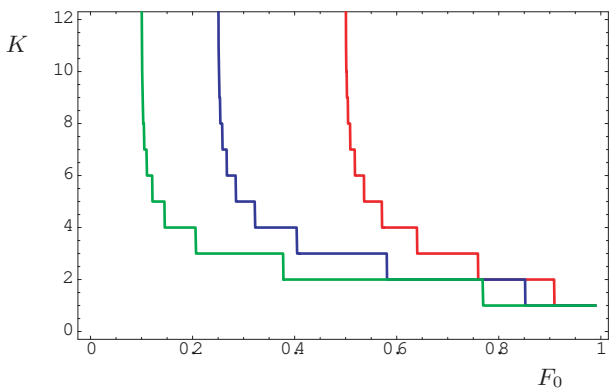}
\caption{Plot of the number of iterations $K(\epsilon, F_0)$ (\ref{g22})
to achieve a final fidelity of $F=0.99$ as a function of the initial
fidelity $F_0$ and for several values of the dimension $D$ of the
qudits: $D=2,4,10$.}
\label{iterations}
\end{figure}
In Fig.~\ref{iterations} we plot the number of iterations (\ref{g22}) for
a given value of the final fidelity $1-\epsilon$ that we take as the fixed
value of $0.99$, and then we find how is the dependence on the initial fidelity
$F_0$. We see that for a given admissible value of $F_0$, the lowest number
of iterations corresponds to the protocol with the higher value of the qudits
dimension $D$:
\begin{equation}
K(\epsilon_0, F_0)_{D_1} \geq K(\epsilon_0, F_0)_{D_2}, \ {\rm for} \ D_1>D_2.
\label{g23}
\end{equation}

\begin{figure}[t]
\psfrag{x}[Bc][Bc][1][0]{$q_0$}
\psfrag{y}[Bc][Bc][1][0]{$q_1$}
\psfrag{z}[Bc][Bc][1][0]{$q'_0$}
\includegraphics[width=8 cm]{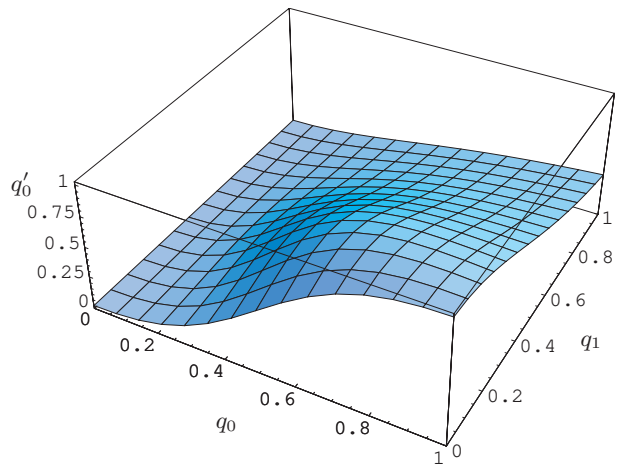}
\caption{The desired fidelity $q'_0$ for amplification after a single
application of the distillation protocol for qutrits (\ref{g25}) as a function
of the previous fidelitites $q_0,q_1$.}
\label{qutrits}
\end{figure}

\begin{figure}[t]
\psfrag{A}[Bc][Bc][1][0]{$(0,0)$}
\psfrag{B}[Bc][Bc][1][0]{$(\half,0)$}
\psfrag{C}[Bc][Bc][1][0]{$(1,0)$}
\psfrag{D}[Bc][Bc][1][0]{$(\half,\half)$}
\psfrag{E}[Bc][Bc][1][0]{$(0,1)$}
\psfrag{F}[Bc][Bc][1][0]{$(0,\half)$}
\psfrag{X}[Bc][Bc][1][0]{$q_0$}
\psfrag{Y}[Bc][Bc][1][0]{$q_1$}
\includegraphics[width=6 cm]{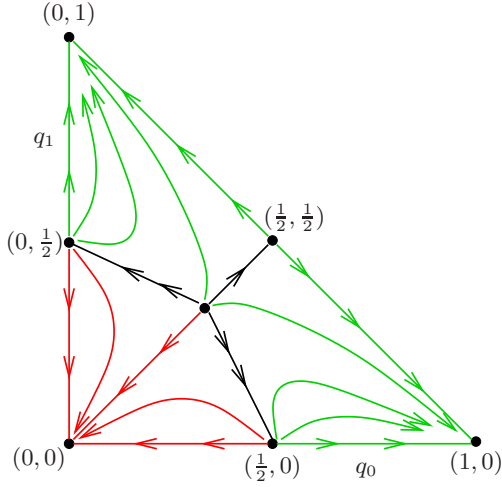}
\caption{Two-dimensional flow diagram associated to the distillation protocol
for qutrits (\ref{g25}).}
\label{2Dflow}
\end{figure}

{\em iii).} For qutrits, $D=3$, the most general diagonal mixed state with the
allowed Bell states taking values on the set $\{ \ket{\Psi_{0i}} \}$ is
\begin{equation}
\begin{split}
\rho &:= q_0\ket{\Psi_{00}} + q_1\ket{\Psi_{01}} + q_2\ket{\Psi_{02}} \\
1 &=: q_0 + q_1 + q_2
\end{split}
\label{g24}
\end{equation}
Let us assume that the state we want to purify is $\ket{\Psi_{00}}$.
Now, our recursion relation for our
fidelity $q_0$ depends on two variables, namely,
\begin{equation}
q'_0 = \frac{q_0^2}{q_0^2 + q_1^2 + (1-q_0-q_1)^2},
\label{g25}
\end{equation}
and a similar equation for $q_1$ with $q_0 \leftrightarrow q_1$.
In Fig.~\ref{qutrits} the dependence of the function fidelity
$q'_0=q'_0(q_0,q_1)$ for qutrits is plotted. We observe that it is
a monotonous incresing function which gurantees that the initial
fidelity will flow towards 1, under certain conditions.
To find these conditions, we find that the set of fixed points
of these recursion relations is given by
$$(q_0,q_1)_c=\{ (0,0), (\half,0), (0,\half), (\frac{1}{3},\frac{1}{3}),
(\half,\half), (1,0), (0,1) \}.$$
We have also found the flow diagram associated to these recursion relations
which is now two-dimensional and we show it in Fig.~\ref{2Dflow}. From this
diagram we see that the purification protocol is successful in arriving to
the maximum fidelity $q_0=1$ provided the initial fidelity lies in the
stability basin of the fixed point $(1,0)$ which is given by the trapezoid
formed by the set of points
$(\half,\half),(\frac{1}{3},\frac{1}{3}),(\half,0),(1,0)$.

\section{Distillation of Non-Diagonal Mixed States}
\label{sec4}
So far, we have been investigating the properties of distillation
protocols applied to mixed bipartite state of diagonal form such
as those in (\ref{g13}), (\ref{g14}). It is apparent that once we have a general result
for the operation of the $U_{\rm BCNOT}$ gate of generalized Bell states
(\ref{g8}), we can also deal with non-diagonal mixed states, namely,
\begin{equation}
\begin{split}
\rho := \sum_{k,j=0}^{D-1}\sum_{k',j'=0}^{D-1}
q_{kjk'j'}\ket{\Psi_{kj}}\bra{\Psi_{k'j'}},\\
1 =: \sum_{k,j=0}^{D-1}\sum_{k',j'=0}^{D-1}
q_{kjk'j'}, \ q_{kjk'j'}\geq 0.
\end{split}
\label{g26}
\end{equation}
Since this is a too much general state, we prefer to extract from this class
of non-diagonal states one type which we believe it may have potential
applications.

Let us imagine that Alice and Bob are manipulating bipartite qudit states
that are {\em diagonal} in the computational basis. More explicitly, the
entangled state they want to purify is of the form
\begin{equation}
\ket{\Psi_{\rm d}} := \frac{1}{\sqrt{D}}\sum_{i=0}^{D-1}\ket{ii},
\label{g27}
\end{equation}
while states which are {\em non-diagonal} are considered as acting as
disturbing noise that they want to get rid of. Specifically, this noise
will be represented by the state
\begin{equation}
\ket{\Psi_{\rm o}} := \frac{1}{\sqrt{D(D-1)}}\sum_{i\neq j=0}^{D-1}\ket{ij}.
\label{g28}
\end{equation}
Then, in order to achieve their goal of purifying states of the diagonal
form $\ket{\Psi_{\rm d}}$ with respect to non-diagonal states
$\ket{\Psi_{\rm o}}$, they set up a distillation protocol based on sharing
copies of the following mixed state
\begin{equation}
\rho := F \ket{\Psi_{\rm d}}\bra{\Psi_{\rm d}}
+ (1-F) \ket{\Psi_{\rm o}}\bra{\Psi_{\rm o}}.
\label{g29}
\end{equation}
We envisage that this scenario is physically feasible since we can imagine
that the computational basis is realized in terms of some physical property
taking values on $i=0,\ldots,D-1$ and that Alice and Bob have a mechanism to
select when they have $\ket{ii}$ coincident qudits (or diagonal) from
$\ket{ij}$ non-coincident qubits (non-diagonal).

To proceed with the distillation of the state $\rho$ in (\ref{g29}),
we first must express the states $\ket{\Psi_{\rm d}}$ and $\ket{\Psi_{\rm o}}$
in the basis of the generalized Bell states, with the result
\begin{equation}
\begin{split}
\ket{\Psi_{\rm d}} &= \ket{\Psi_{00}},\\
\ket{\Psi_{\rm o}} &= \frac{1}{\sqrt{D-1}} \sum_{i=1}^{D-1} \ket{\Psi_{0i}}.
\end{split}
\label{g30}
\end{equation}
Next, Alice and Bob share two pairs of non-diagonal mixed states
\begin{equation}
\begin{split}
&\rho \otimes \rho = F^2 \ket{\Psi_{00} \Psi_{00}}\bra{\Psi_{00} \Psi_{00}} \\
&+ \frac{F(1-F)}{D-1}\sum_{i,j=1}^{D-1}
[\ket{\Psi_{00} \Psi_{0i}}\bra{\Psi_{00} \Psi_{0j}} +
\ket{\Psi_{0i} \Psi_{00}}\bra{\Psi_{0j} \Psi_{00}}]\\
&+ \frac{(1-F)^2}{(D-1)^2}\sum_{i,j,k,l=1}^{D-1}
\ket{\Psi_{0i} \Psi_{0k}}\bra{\Psi_{0j} \Psi_{0l}},
\end{split}
\label{g31}
\end{equation}
and they apply bilaterally the CNOT gate to it (\ref{g8}) with the result
\begin{equation}
\begin{split}
&U_{\rm BCNOT}\rho \otimes \rho U_{\rm BCNOT} = F^2 \ket{\Psi_{00} \Psi_{00}}\bra{\Psi_{00} \Psi_{00}} \\
&+ \frac{F(1-F)}{D-1}\sum_{i,j=1}^{D-1}
[\ket{\Psi_{00} \Psi_{0,\ominus i}}\bra{\Psi_{00} \Psi_{0,\ominus j}} +
\ket{\Psi_{0i} \Psi_{0i}}\bra{\Psi_{0j} \Psi_{0j}}]\\
&+ \frac{(1-F)^2}{(D-1)^2}\sum_{i,j,k,l=1}^{D-1}
\ket{\Psi_{0i} \Psi_{0i\ominus k}}\bra{\Psi_{0j} \Psi_{0j\ominus l}}.
\end{split}
\label{g32}
\end{equation}
The process of measuring the target qudits and retaining the
source qudits when upon classical communication Alice and Bob
find coincidences in their measures amounts to retaining the terms in
(\ref{g32}) that have the state $\ket{\Psi_{00}}$ in the target qudits.
This means that only the first term and part of the last term in (\ref{g32})
contribute to the final source mixed state, which takes the following
form without normalization
\begin{equation}
\begin{split}
\rho'  &\sim F^2 \ket{\Psi_{00}} \bra{\Psi_{00}}
+ \frac{(1-F)^2}{(D-1)^2}\left(\sum_{i=1}^{D-1} \ket{\Psi_{0i}}\right)
\left(\sum_{j=1}^{D-1} \bra{\Psi_{0j}}\right) \\
&= F^2 \ket{\Psi_{00}} \bra{\Psi_{00}} + \frac{(1-F)^2}{D-1}
\ket{\Psi_{\rm o}} \bra{\Psi_{\rm o}}.
\end{split}
\label{g33}
\end{equation}
Upon normalization, we arrive again at a non-diagonal mixed state of
same form as the one we started with
$\rho' = F' \ket{\Psi_{\rm d}}\bra{\Psi_{\rm d}}
+ (1-F') \ket{\Psi_{\rm o}}\bra{\Psi_{\rm o}}$,
but with a new fidelity $F'$ given by
\begin{equation}
F'= \frac{F^2}{F^2 + \frac{(1-F)^2}{D-1}}.
\label{g34}
\end{equation}
Let us notice that this is precisely the same recursion relation that
we found in Sect. III in a different context (\ref{g18}).

\section{Continuum Limit of Qudit Protocols}
\label{sec5}
For the general case represented by the recursion relations (\ref{g16})
we can also find the general solution for the $k$-th iteration
$q_i^{(k)}, i=0,\ldots,D-1$
starting from their initial values $q_i^{(0)}$ satisfying
$\sum_{i=0}^{D-1} q_i^{(0)}:=1$. We find the following solution
\begin{equation}
q_i^{(k)} = \frac{\left[q_i^{(0)}\right]^{2^k}}
{\sum_{j=0}^M \left[q_j^{(0)}\right]^{2^k}}.
\label{g16b}
\end{equation}
Let us assume that the maximum initial value is $M:={\rm max}\ \{ q_i^{(0)}\}$
and it is $p$ times degenerate. Then, using the general solution (\ref{g16b})
we can inmeditely find the fixed points after the evolution with the recursion
relations. We find
\begin{equation}
\begin{split}
\lim_{k\rightarrow \infty} q_i^{(k)} = 0, \ {\rm if} \ q_i^{(0)}<M,\\
\lim_{k\rightarrow \infty} q_i^{(k)} = \frac{1}{p}, \ {\rm if} \ q_i^{(0)}=M.
\end{split}
\label{g16c}
\end{equation}

From the analysis of these distillation protocols and the way they operate
we arrive at the conclusion that they resemble a sort of
amplitude amplification quite similar to what happens in the Grover algorithm
where there exists what is called {\em quantum amplitude amplification}.
However, there is an important distintion between both procedures:
in the distillation method, the maximum amplification is attained
asymptotically, while in Grover algorithm it is achieved periodically.
The reason for this difference relies on the fact that the distillation process
is not unitary (since we make measurements and discard states), while Grover
is unitary. Thus, we propose to refer to the distillation protocol as
{\em weight amplification}, since it is certain probability weights of the
intial mixed states, and not amplitudes, what are being amplified.

When $D$ is very large, we can approximate the probability weights
$q_i^{(0)}$ taking values on the discrete set $\{ 0,1,\ldots,D-1\}\ni i$,
by a density function $q(x)^{(0)}$ defined on the real interval $[0,1]$.
This is achieved by introducing the variable $x\in[0,1]$ defined as
$x:=i\Delta x$ with $\Delta x:=\frac{1}{D-1}$. Thus, in the limit
$D\rightarrow \infty (\Delta x \rightarrow 0)$, we get a probability density
as $q_i^{(0)}(i\Delta x)\rightarrow q^{(0)}(x)dx$. It is also normalized as

\begin{equation}
\int_{0}^1 q^{(0)}(x) dx = 1.
\label{g35}
\end{equation}
Likewise, we can take the continuum limit of the general recursion equation
(\ref{g16b}) in order to obtain the probability density $q(x)^{(k)}$
after $k$ steps of the distillation protocol. This is given by
\begin{equation}
q^{(k)}(x) = \frac{\left[q^{(0)}(x)\right]^{2^k}}
{\int_{0}^1 \left[q^{(0)}(y)\right]^{2^k} dy}.
\label{g36}
\end{equation}
\begin{figure}[t]
\psfrag{x}[Bc][Bc][1][0]{$x$}
\psfrag{y}[Bc][Bc][1][0]{$q^{(k)}(x)$}
\vspace{10 pt}\includegraphics[width=8 cm]{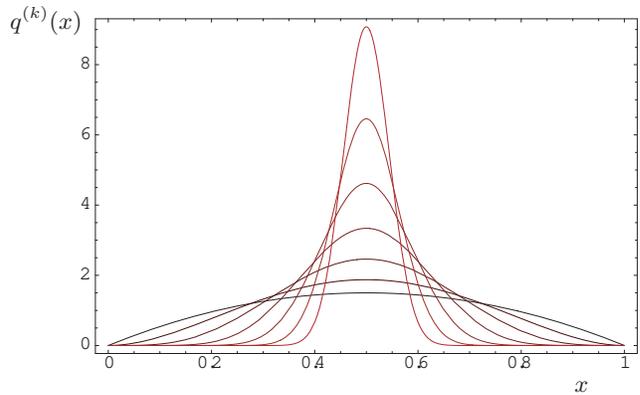}
\caption{Evolution of the probability density under
the iterative application of the distillation protocol in the
continuum limit (\ref{g36}). Starting with a parabolic
distribution $k=0$, we show the resulting profiles after steps
$k=1,2,\ldots,6$.} \label{evolution}
\end{figure}
This is a closed analytical equation that provides us with the evolution
of the probability density for any initial probability profile $q^{(0)}(x)$.
In Fig.~\ref{evolution} we plot this evolution for an initial distribution
of a parabolic form $q^{(0)}(x) = 6(x-x^2)$. We see how as we increase the
step $k$ of the distillation, the new distributions get peaked around the
highest value of the initial distribution, which is $x=\half$ in this
particular case. This behaviour illustrates the idea of the
weight amplification and is in agreement with the results
(\ref{g16c}) for the fixed points of the flow equations.

\section{Quantum Distillation and Quantum Renormalization}
\label{sec6}
\begin{table*}[Ht]
\begin{ruledtabular}
\begin{tabular}{cc}
{\bf Quantum Distillation} & {\bf Quantum RG}  \\
\hline \hline
Mixed State $\rho$ & Quantum Hamiltonian $H$ \\ \hline
Computational Basis & Local Site Basis \\ \hline
Bell Basis & Energy Basis \\ \hline
Alice \& Bob Tensor Product & Blocking Method \\ \hline
L.O.C.C. & Truncation Operator \\ \hline
Maximum Fidelity & Minimum Energy \\ \hline
RG-Flow Diagram & Distillation-Flow Diagram \\
\end{tabular}
\end{ruledtabular}
\caption{Summary of the comparative analysis between the quantum distillation
process and a quantum renormalization group method for lattice Hamiltonians.}
\label{tableQDQRG}
\end{table*}

It is interesting to notice the analogy between the recursive distillation
process represented by the equation (\ref{s6}) and Fig.~\ref{flowdiagram}
and the truncation process in the Renormalization
Group analysis of certain quantum lattice Hamiltonian models, specifically,
the ITF model (Ising in a Transverse Field) \cite{jaitisi},\cite{analytic}.
The basic idea of a QRG method is: i/ elimination of high energy states plus,
ii/ iterative process. This is precisely what happens in a quantum distillation
process which we have seen in the preceeding sections, achieving a purification of a mixed state by means of discarding states and a recursive procedure.
This relationship can be made even closer if we briefly recall what a
quantum renormalization group (QRG) method is.
The subject of the distillation is a mixed state operator $\rho$, while that
of the renormalization is a quantum Hamiltonian operator $H$.
A summary of these relations
is presented in Table~\ref{tableQDQRG} that will be deduced along the way.
\begin{figure}\begin{center}
\includegraphics[scale=0.35]{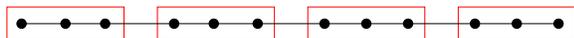}
\caption{Block decomposition of the Heisenberg chain in 3-site blocks.}
\label{chainbrg}
\end{center}\end{figure}
\begin{figure}[h]
\begin{center}
\psfrag{0}[Bc][Bc][0.751][0]{$F_0$}
\psfrag{1}[Bc][Bc][0.751][0]{$F_1$}
\psfrag{2}[Bc][Bc][0.751][0]{$F_2$}
\psfrag{3}[Bc][Bc][0.751][0]{$F_3$}
\includegraphics[width=9 cm]{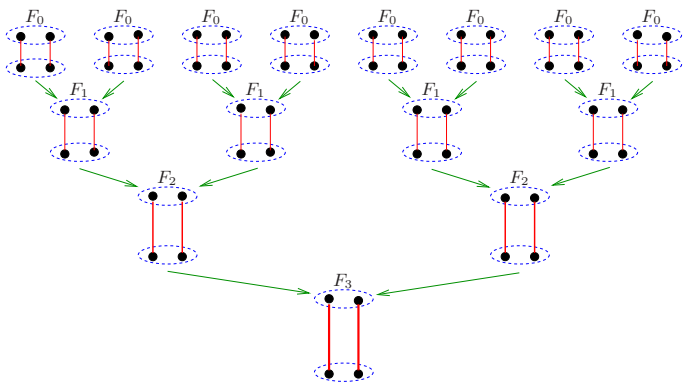}
\caption{Example of distillation process starting with 8 pairs of mixed
states for Alice (and the same amount for Bob). After 3 steps the original
fidelity $F_0$ is improved up to a final value of $F_3$ (we assume full
success for simplicity).}
\label{binarydistill}
\end{center}
\end{figure}
\begin{figure}
\begin{center}
\psfrag{c}[Bc][Bc][1][0]{Initial chain with 8 sites}
\psfrag{R}[Bc][Bc][1][0]{ QRG}
\includegraphics[width=8 cm]{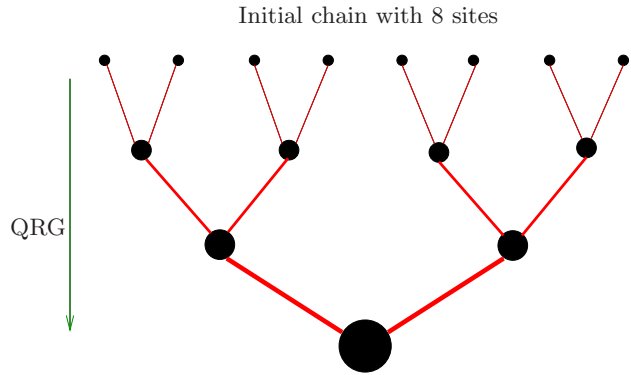}
\caption{Example of renormalization QRG process for an initial chain
with 8 sites in 3 steps (making blocks of 2 sites each).}
\label{binaryflow}
\end{center}
\end{figure}
The easiest way to present the QRG method  is with an example of
quantum lattice Hamiltonian like the isotropic Heisenberg model on a 1D chain:
\begin{equation}
H = J \sum_{i=0}^N \vec{S}_i\cdot \vec{S}_{i+1},
\label{blocking12}
\end{equation}
with $\vec{S}_i$ spin-$\half$ operators at site $i$ of the chain.
The local site basis $\{ \ket{\!\downarrow}, \ket{\!\uparrow} \}$
corresponds to
the computational basis $\{\ket{0}, \ket{1}\}$. Much like this latter basis is
not enough for doing the distillation, the local site basis needs to be
complemented with another type of basis. To see this, let us start the RG
process with the block decomposition of the chain in blocks of
$n_B=3$ sites as shown in Fig.~\ref{chainbrg}. This blocking method in QRG
corresponds to the tensor product of Alice and Bob's shared states at the
begining of the distillation process, as shown in Fig.~\ref{binarydistill}.
This is to be compared with the similar iterative process in the QRG method
in Fig.~\ref{binaryflow}.

The block Hamiltonian is then
\begin{equation}
\begin{split}
H_{B} &=  J (\vec{S}_1\cdot \vec{S}_{2} + \vec{S}_2\cdot \vec{S}_{3}) \\
&=\frac{{ J}}{2} \left[ (\vec{S}_1 + \vec{S}_{2} + \vec{S}_{3})^2 -
\vec{S}_{2}^2 - (\vec{S}_1 + \vec{S}_{3})^2\right]
\end{split}
\label{blocking13}
\end{equation}
The label $B$ here stands for Block and not for Bob.
The diagonalization of $H_{B}$ is straightforward using the
Clebsch-Gordan decomposition of the tensor product of 3 irreducible representations of spin $S=\half$,
\begin{equation}
\half { \otimes} \half { \otimes} \half = \half { \oplus} \half { \oplus} \frac{3}{2}.
\label{blocking14}
\end{equation}
In particular, the ground state (GS) is given by
\begin{equation}
\ket{\!\Uparrow}_{\rm GS} = \frac{1}{\sqrt{6}}\left[ 2\ket{\uparrow \downarrow \uparrow}
- \ket{\downarrow \uparrow \uparrow} - \ket{\uparrow \uparrow \downarrow}
\right],
\label{blocking15}
\end{equation}
which is a spin doublet (with a similar expresion for the other state
$\ket{\!\Downarrow}_{\rm GS}$, with the spins
reversed). This fact is peculiar of the 3-site block and it is
the main underlying reason for using a block of that size in the QRG (this fact
is model dependent: for the ITF model, the blocking is with $n_B=2$ sites
\cite{jaitisi},\cite{analytic}, Fig.~\ref{binaryflow}).
In the energy basis, the block Hamiltonian is diagonal and this corresponds
to the Bell basis for the mixed state $\rho$ in the distillation process.

\begin{figure}\begin{center}
\includegraphics[scale=0.45]{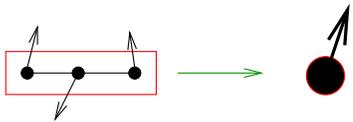}
\caption{QRG renormalization of $H_B$.}
\label{chain3spins}
\end{center}\end{figure}

Now, the truncation of states amounts to retaining the state of lowest energy
(doublet) and discarding the remaining 2 excited states. This reduction
scheme is of the form $2^3=8\longrightarrow 2$. This truncation corresponds
to discarding unwanted states of non-coincidences in the distillation process.
The new effective site
is again a spin-$\half$ site as shown in Fig.~\ref{chain3spins}.
The RG-truncation is implemented by means of a truncation operator $O$
constructed from the lowest energy eigenvalues of $H_B$ retained during the
renormalization process. In this example, $O$ is constructed from the lowest
energy doublet in the Clebsch-Gordan decomposition (\ref{blocking14}),
namely (\ref{blocking15}).
Similarly, in the distillation process we have that the tensor product of
Alice and Bob's states can be decomposed into states with coincident qubits
in the target, denoted by $\rho^{a}_{\rm C}$ (\ref{s3}), and states with non-coincident qubits in the target, denoted by $\rho^{a}_{\rm NC}$
(\ref{s3}), i.e.,
\begin{equation}
\rho_{AB}\otimes \rho_{AB} =
(\sum_{a} \rho^{a}_{\rm C})\oplus (\sum_{a} \rho^{a}_{\rm NC}),
\label{distill14}
\end{equation}
where the sum in $a$ runs over a certain number of mixed states of 4 parties.
Notice that this stage is similar to the RG-stage
represented by equation (\ref{blocking14}).
Next, an elimination process similar to
the RG-truncation is performed by means of LOCC operations (measurements and
classical communication) that retains only the bipartite states embedded in
the $\rho^{a}_{\rm C}$ states.

Then, the renormalization of the block Hamiltonian is simply
\begin{equation}
{ H_{B'}} = { O} { H_B} { O^{\dagger}} = E_0 = -{ J}.
\label{blocking16}
\end{equation}
Similarly, we could have expressed the state-elimination of the distillation
in previous sections in terms of an truncation operator, say $O_D$, such that
the new mixed state $\rho'_{AB}$ is obtained as
\begin{equation}
{\rho'_{AB}} = { O_D} (\rho_{AB}\otimes \rho_{AB}) { O^{\dagger}_D}.
\label{distill16}
\end{equation}

\begin{figure}\begin{center}
\psfrag{c}[Bc][Bc][1][0]{$\vec{S}_r$}
\psfrag{d}[Bc][Bc][1][0]{$\vec{S}_l$}
\psfrag{r}[Bc][Bc][0.75][0]{r}
\psfrag{l}[Bc][Bc][0.75][0]{l}
\psfrag{a}[Bc][Bc][1][0]{$n$}
\psfrag{b}[Bc][Bc][1][0]{$n+1$}
\includegraphics[scale=0.45]{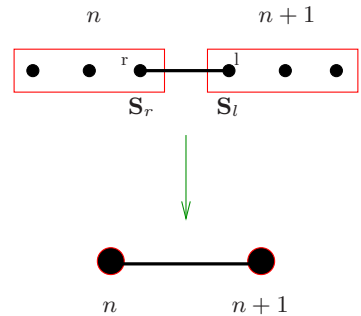}
\caption{QRG renormalization of the interblock Hamiltonian $H_{BB}$.}
\label{chainrlblocks}
\end{center}\end{figure}

In the case of the quantum Hamiltonian, we still need extra work since there
are interaction links between blocks (see Fig.~\ref{chainbrg}).
These are absent in
the distillation protocol. However, the renormalization of the interblock
Hamiltonian $H_{BB}$ follows also the same prescription as in
(\ref{blocking16}) and we arrive at
\begin{equation}
{  J} \vec{S}_r^{n} \cdot \vec{S}_l^{n+1} \xrightarrow{\rm {  RG}}
{  J} (\frac{2}{3})^2 \vec{S}'_{n} \cdot \vec{S}'_{n+1},
\label{blocking18}
\end{equation}
where we denote by $n$ and $n+1$ two successive blocks in the original lattice
(see Fig.~\ref{chainrlblocks}) that become two successive sites
(see Fig.~\ref{chain3spins})
in the new lattice after the renormalization. We can collect all these steps
in Table~\ref{tableQRG} \cite{white}, \cite{jaitisi}. This table should be
contrasted with the similar table for the distillation process that can be
formed with the steps explained in Sect.~\ref{sec2}.

\begin{table}
\begin{ruledtabular}
\begin{tabular}{l}
1/ Block Decomposition: $H=H_B+H_{BB}$. \\ \hline
2/ Diagonalization of $H_B$. \\ \hline
3/ Truncation within each Block: $O$. \\ \hline
4/ Renormalization: $H'_B= O H_B O^{\dagger}$,
$H'_{BB}= O H_{BB} O^{\dagger}$ \\ \hline
5/ Iteration: Go to 1/ with $H'=H'_B+H'_{BB}$.\\
\end{tabular}
\end{ruledtabular}
\caption{Steps of the quantum renormalization group method (QRG)
for lattice Hamiltonians.}
\label{tableQRG}
\end{table}

The outcome of the RG-method is that we obtain the correct
RG-flow for the coupling constant $J\longrightarrow 0$, signalling a gapless
system plus an approximate estimation for the ground state energy, which
by means of the variational principle, it is an upper bound for the exact
energy. Therefore, the QRG is an energy minimization procedure.
Likewise, the purification process produces a protocol for
fidelity maximization along with a distillation-flow diagram.
This completes the relationship established in
Table~\ref{tableQDQRG} between quantum
distillation and quantum renormalization.

\section{Conclusions}
\label{sec7}

The field of quantum distillation protocols  has become very active
in the theory of quantum information due to the central role played
by entanglement in the quantum communication procedures and its tendency
to degradation.

In this work we have been interested in several extensions of the purification
protocols when dealing with multilevel systems (qudits) instead of the
more usual qubit protocols. We have seen the various advantages of having
distillation methods for qudits systems as compared with the simple case of
qubits. We have also obtained the general form of the solution to the
distillation recursion relations and several particular solutions have been
studied explicitly. We have developed the relationship between
quantum distillation protocols and quantum renormalization group methods,
something which is interesting in itself and could serve as a guide
for possible extension of purification methods.

We would like to mention that the possibility of working with qudits systems
has become quite realistic in the recent years. For instance, it is possible to
realize multilevel systems in terms of the orbital angular momentum of
photons, instead of the more standard polarization (qubit) degree of freedom
\cite{zeilinger2}, \cite{arnaut}, \cite{barnett}. Yet another possibility
is to use the so called multiport beam splitters
\cite{reck}, \cite{zukowski}, \cite{molina-terriza}, \cite{padgett}.

There are several ways in which this work can be extended. One is the
consideration of noise as as source of errors during the distillation
protocol itself. Another one is to allow the possibility of having these
distillation protocols for qudits be
embedded into a quantum repeater protocol \cite{repeaters-pur},
\cite{repeaters-com}.

\noindent {\em Acknowledgments}. This work is partially supported by the
DGES under contract BFM2000-1320-C02-01.

\appendix

\section{General Solution of the Distillation Recursion Relations}
\label{app}
In this appendix, we look for more general solutions to the general
distillation recursion relations (\ref{g14}) than those studied in
section~\ref{sec3}.
To this end, it is convenient to introduce auxiliary variables $g_{kj}^{(n)}$
defined by
\begin{equation}
g_{kj}^{(n)} = \sum_{k'=0}^{D-1} g_{k\ominus k'j}^{(n-1)} g_{k'j}^{(n-1)},\
g_{kj}^{(0)} = q_{kj}^{(0)},
\label{ap1}
\end{equation}
so that the real weights $q_{kj}^{(n)}$ are related to these auxiliary variables as
\begin{equation}
q_{kj}^{(n)} := \frac{g_{kj}^{(n)}}{\sum_{l,i=0}^{D-1} g_{li}^{(n)}}.
\label{ap2}
\end{equation}
Thus, $g_{kj}^{(n)}$ are unnormalized probability weights. The recursion
relations they satisfy can be read as follows (\ref{ap1}): for a fixed
second index $j$, the unnormalized weights $g_{kj}^{(n)}$ at the step $n$
of the distillation process are obtained as the convolution over the first
indices $k$ of the unnormalized weights $g_{kj}^{(n-1)}$ in an earlier step.
This fact calls for the introduction of the Fourier transform in order to
analyze the relations (\ref{ap1}). Let us introduce the new variables
$R_j^{(n)}$ defined as
\begin{equation}
R_{\hat{k}j}^{(n)} := \sum_{k=0}^{D-1} \ee^{\frac{2\pi \ii  \hat{k} k}{D}} g_{kj}^{(n)}.
\label{ap3}
\end{equation}
Now, using the properties of the convolution and the Fourier transform it is
inmediate to arrive at a simpler recursion relation
\begin{equation}
R_{\hat{k}j}^{(n)} = \left[R_{\hat{k}j}^{(n-1)}\right]^2,
\label{ap4}
\end{equation}
which can be iterated all the way down to the initial step
\begin{equation}
R_{\hat{k}j}^{(n)} = \left[R_{\hat{k}j}^{(0)}\right]^{(2^n)}.
\label{ap5}
\end{equation}
Fourier transforming back to the unnormalized variables, we get
\begin{equation}
g_{kj}^{(n)} = \frac{1}{D}\sum_{\hat{k}=0}^{D-1}
\ee^{-\frac{2\pi \ii k \hat{k}}{D}}
\left[ \sum_{k'=0}^{D-1} \ee^{\frac{2\pi \ii \hat{k}k'}{D}} q_{k'j}^{(0)}
\right]^{(2^n)},
\label{ap6}
\end{equation}
from which we also obtain the normalized probability weights $q_{kj}^{(n)}$
upon normalization (\ref{ap2}). In particular, for the case of qubits treated
in Sect.~\ref{sec2}, $D=2$ and if we also restrict ourselves
 to weights of the form 
$q_{00}=F, q_{01}=1-F, q_{10}=q_{11}=0$, we again 
obtain from the general solution
(\ref{ap6}) the simple recursion relation in equation (\ref{s6}).


\end{document}